\documentclass[aps,pre,twocolumn,groupedaddress]{revtex4-1}

\usepackage{graphicx}
\usepackage{amsmath}

\begin{document}


\title{Eigenvector dynamics under perturbation of modular networks}


\author{Somwrita Sarkar}
\email{somwrita.sarkar@sydney.edu.au}
\thanks{corresponding author}
\affiliation{Design Lab, Faculty of Architecture, Design and Planning, University of Sydney, Australia NSW 2006}
\affiliation{ARC Centre of Excellence for Integrative Brain Function}

\author{Sanjay Chawla}
\affiliation{Faculty of Engineering and IT, University of Sydney, Australia NSW 2006}

\author{P. A. Robinson}
\affiliation{School of Physics, University of Sydney, Australia NSW 2006} 
\affiliation{ARC Centre of Excellence for Integrative Brain Function}

\author{Santo Fortunato}
\affiliation{Department of Computer Science, Aalto University, Finland}


\date{\today}

\begin{abstract}
Rotation dynamics of eigenvectors of modular network adjacency matrices under random perturbations are presented. In the presence of $q$ communities, the number of eigenvectors corresponding to the $q$ largest eigenvalues form a ``community" eigenspace and rotate together, but separately from that of the ``bulk" eigenspace spanned by all the other eigenvectors. Using this property, the number of modules or clusters in a network can be  estimated in an algorithm-independent way. A general argument and derivation for the theoretical detectability limit for sparse modular networks with $q$ communities is presented, beyond which modularity persists in the system but cannot be detected. It is shown that for detecting the clusters or modules using the adjacency matrix, there is a ``band" in which it is hard to detect the clusters even before the theoretical detectability limit is reached, and for which the theoretically predicted detectability limit forms the sufficient upper bound. Analytic estimations of these bounds are presented, and empirically demonstrated. 

\end{abstract}


\maketitle

\section{Introduction}
\label{introduction}

Networks have community structure, i. e., groups of nodes with significantly higher internal link density than the density of links joining the groups. Community detection, the problem of correctly estimating the number of communities and their constitution, has attracted significant attention in physics, applied mathematics and computer science~\cite{newman2010, fortunato2010}. Accurate solutions enhance the understanding of the relationships between network structure and dynamics.

Spectral methods, employing the eigenvectors and eigenvalues of the adjacency, Laplacian, and modularity matrices~\cite{newman06,newman06b,fortunato2010, newman2010, sarkar2011a, sarkar2013a, krzakala2013,newman13}, are widely used to identify communities. While the behaviour of eigenvalues is widely studied~\cite{fortunato2010, newman2010, nadakuditi2012, sarkar2011a, sarkar2013a, chauhan2009, krzakala2013}, there is less work on understanding how eigenvectors behave under variations in network structure, even though it is the eigenvector properties that are used to perform community detection. In the present paper, the focus is on the behaviour of the eigenvectors of the adjacency matrix and the relationship of this behaviour to gaps between eigenvalues of the adjacency matrix. It is to be expected that similar results will hold for Laplacian and modularity matrices also. 

A related problem is the algorithm-independent determination of the number of communities, a parameter that many detection methods need as input. Other methods estimate this number, but several runs of the same algorithm on even the same data set can return different numbers and constitutions of communities. The performance of several algorithms in determining this number has been measured~\cite{darst2014}, and its a priori knowledge improves their performance significantly. Algorithm-independent techniques and analytic understanding of systems to determine this number are thus beneficial. 

One algorithm-independent way of determining the number of modules is to count the number of eigenvalues $q$ separated from the bulk eigenvalues of a suitable matrix representation~\cite{chauhan2009, nadakuditi2012, krzakala2013,newman13b}. However, for networks with broad distributions of node degree, and numbers and sizes of communities, the eigenvalues can show highly variable behaviour. For example, large eigenvalues can reflect both high degree nodes as well as the number of modules. Further, as mentioned above, even though the formal identification of modules is performed based on the properties of the corresponding eigenvectors~\cite{newman2010, fortunato2010, sarkar2011a}, the overall behaviour of eigenvectors under variations in network structure is much less understood than that of the eigenvalues~\cite{newman2010, fortunato2010, nadakuditi2012, rourke2013}. Therefore, this warrants further attention onto the structure of eigenvectors.   
\begin{figure*}
\includegraphics[scale = 0.16]{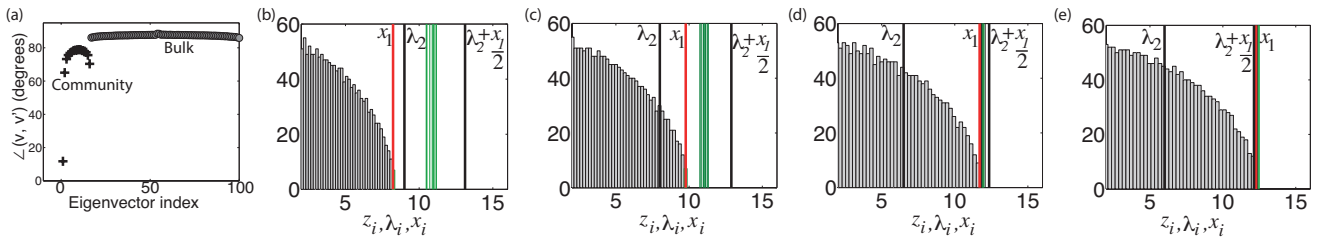}
\caption{\label{Fig1} [Color online] Eigenvector rotation angles under perturbation and their relationship to eigenvalue gaps. (a) Mean angles of rotation under perturbation for eigenvectors corresponding to the 100 largest (absolute) eigenvalues, for stochastic block models with $N = 10000$, $q = 16$, $\mu_{in} = 10$, $\mu_{out} = 1$. Points are averages over $10$ networks, each perturbed $300$ times. (b)-(e) Spectrum for $A$ showing bulk eigenvalues, red line shows $x_{1}$, green lines are $z_{2:q}$ and black lines show their lower bound [$b_{l}$] and upper bound [$b$], $N=4096$, $q=8$, $\mu_{in} = 10$, and $\mu_{out} = 1$, $2$, $3.5$ (just before the detectability limit), and $4$ (just after the detectability limit), respectively.}
\end{figure*}

\vspace{-0.5cm}
\subsection{Contributions}

We investigate rotations of eigenvectors of the adjacency matrix when the network is randomly perturbed: this rotation behaviour is dependent on the gaps between eigenvalues of the adjacency matrix and contains accurate community structure information. The first main result of the paper is that in the presence of $q$ communities, the number of eigenvectors corresponding to the $q$ largest eigenvalues form a ``community" eigenspace and rotate together, but separately from that of the ``bulk" eigenspace spanned by all the other eigenvectors. Using this property, the number of modules or clusters in a network can be  estimated in an algorithm-independent way. We investigate this behaviour right to the theoretical detectability limit, beyond which modularity persists in the system but cannot be detected~\cite{nadakuditi2012, krzakala2013}. The second contribution of the paper is that we present a general derivation of the theoretical detectability limit for $q$ communities, using arguments about upper and lower bounds on the eigenvalues, that was previously shown for the $q=2$ case~\cite{nadakuditi2012}. Third, again using the same bounds, we show that for detecting the clusters or modules using the adjacency matrix, there is a ``band" in which it is hard to detect the clusters even before the theoretical detectability limit is reached, and for which the theoretically predicted detectability limit forms the upper bound. Analytic estimations of these bounds are presented, and empirically demonstrated.   

\section{Background}
\label{background}

A symmetric adjacency matrix $A$ represents an undirected graph $G$ with $N$ nodes, with $A_{ij} = 1$ if an edge exists between nodes $i$ and $j$, and $0$ otherwise. A random Bernoulli perturbation $E$, by definition, is a matrix with half of its entries set to $+1$ and half to $-1$. $E$ can be either symmetric or asymmetric, but for this paper, we assume a symmetric form: since $A$ is symmetric (undirected graph), so we would like $A+E$ to be symmetric. Since $A$ is a simple undirected graph (i.e., no self-loops), we also have the diagonal of $E$ set to 0. $E$ is scaled by a small number $\epsilon$ to control the size of the perturbation, and we construct the perturbed matrix $A+\epsilon E$. In practical implementations, we have used a range of values for $\epsilon$, varying it from 0.01 to 0.2 (the figures show results at $\epsilon = 0.05$, for example). The only care to take while choosing $\epsilon$ would be that the noise should not be so large as to override the signal. With this condition satisfied, these results hold for any chosen $\epsilon$. We study the rotation of eigenvectors under perturbation; i.e., the angles between an eigenvector of $A$ and the corresponding one in $A+\epsilon E$.


The eigenvalues of $A$, are arranged as $z_{1} \geq z_{2} \geq \ldots \geq z_{N}$ to define gaps $\Delta_{i} = z_{i} - z_{i+1}$. The Davis-Kahan-Wedin theorem~\cite{daviskahan1969, vu2011, rourke2013} imposes an upper bound on the sine of the angle between $v_{1}$ and $v'_{1}$\begin{equation} 
\label{eq1} 
\sin\angle(v_{1},v'_{1}) \leq \frac{2 \epsilon ||E||}{\Delta_{1}}, 
\end{equation} where $v_{1}$ and $v'_{1}$ are the first eigenvectors of the original and perturbed matrices respectively, and 
$||E||$ is the spectral norm. When $\Delta_{1} \leq 2 \epsilon ||E||$, the theorem is trivially true, as the sine function is bounded above by 1. Thus, $\Delta_{1} > 2 \epsilon ||E||$ for all non-trivial results. 
If $E$ is symmetric, with mean 0 and unit variance, $||E||=2\sqrt{N}$~\cite{furedi1981}. 


A critical point to note before moving ahead is that the behaviour of the angle between the eigenvectors of $A$ and those of $A + \epsilon E$ could be discontinuous and are dependent principally on the gaps between eigenvalues of $A$. Consider this small example~\cite{vu2011}. Let \begin{equation} A = \begin{bmatrix} 
1+\epsilon & 0 \\
0 & 1- \epsilon 
\end{bmatrix} \end{equation} and let \begin{equation} \epsilon E = \begin{bmatrix} 
-\epsilon & \epsilon \\
\epsilon & \epsilon 
\end{bmatrix}, \end{equation} in which case we get \begin{equation} A + \epsilon E = \begin{bmatrix} 
1 & \epsilon \\
\epsilon & 1 
\end{bmatrix}. \end{equation} Now, it is easy to see that while the eigenvalues of $A$ and $A + \epsilon E$ are the same, $1 + \epsilon$ and $1 - \epsilon$, the eigenvectors of $A$ are $[0,1]$ and $[1,0]$, but the eigenvectors of $A + \epsilon E$ are $[1/ \sqrt{2}, 1/ \sqrt{2}]$ and $[1/ \sqrt{2}, -1/ \sqrt{2}]$, regardless of how small $\epsilon$ is. Thus, it turns out that the behaviour of the eigenvectors under perturbation, and the identification of the number of communities, are dependent on the gaps between the eigenvalues. If these gaps are very small, the rotations could be discontinuous and large. 

Recently these bounds were improved for matrices of low rank~\cite{vu2011, vu2011a, rourke2013}. The intuition is that if $A$ has low rank structure, the action of $E$ on $A$ will also occur in a lower rank subspace. Thus, $||E||= O(\sqrt{N})$~[Eq.~(\ref{eq1})] can be replaced by a dependence on the rank $q$ of $A$ because $q < O(\sqrt{N})$, leading to tighter bounds on the rotation of eigenvectors as measured by the sine. 
If a network has $q$ communities, a lower rank matrix of rank $q$ is a suitable representation of the original network matrix. 
The improvements in~\cite{vu2011,vu2011a, rourke2013} show that with high probability,
\begin{equation} 
\label{eq3} 
\sin \angle (v_{1}, v'_{1}) \leq C_{0} \left(\frac{\sqrt{q}}{\Delta_{1}} + \frac{\epsilon||E||}{z_{1}} + \frac{\epsilon^{2}||E||^{2}}{z_{1} \Delta_{1}} \right). 
\end{equation} 
The improvements also provide a bound on the largest principal angles between subspaces $V = \{v_{1}, \ldots, v_{j}\}$ and $V' = \{v'_{1}, \ldots, v'_{j}\}$, for $1 \leq j \leq q$, defined as
\begin{equation} 
\label{eq4a} 
\sin \angle(V,V') = \smash{\displaystyle\max_{v \in V; v \neq 0}}  \, \, \smash{\displaystyle\min_{v' \in V'; v' \neq 0}} \sin \angle(v,v'). 
\end{equation} 
The bound on subspaces is given by 
\begin{equation} 
\label{eq4b} 
\sin \angle (V, V') \leq C_{1} \left(\frac{\sqrt{q}}{\Delta_{j}} + \frac{\epsilon||E||}{z_{j}} + \frac{\epsilon^{2}||E||^{2}}{z_{j} \Delta_{j}} \right). 
\end{equation}
These improvements~\cite{vu2011, vu2011a, rourke2013} provide a tighter bound on the angles than the Davis-Kahan bounds. 



\section{Results}

We construct $A$ using a stochastic block model (SBM), following~\cite{sarkar2013a, sarkar2013b, nadakuditi2012}, with $q$ communities of $s$ nodes, yielding a total number of nodes $N=sq$. Each node $i$ has a community label $g_{i} \in [1, \ldots, q]$. Edges are then generated independently based on a $q \times q$ probability matrix $p$, with $Pr[A_{ij}=1]=p_{g_{i}g_{j}}$. 
In the simplest case, $p_{ab} = p_{in}$ if $a=b$ and $p_{ab}=p_{out}$ if $a\neq b$, with $p_{in}>p_{out}$. For the sparse case, we define $c_{in}=Np_{in}$ and $c_{out}=Np_{out}$, or equivalently $\mu_{in} = sp_{in}$ and $\mu_{out}=sp_{out}$, with $c_{in}$ and $c_{out}$ constant in the limit $N\to\infty$. Thus, $A$ is partitioned into $q^2$ blocks of size $s\times s$, with $q$ blocks along the diagonal and $q(q-1)$ off-diagonal. 

We have empirically shown distributions of eigenvalues and resulting detection of modularity for a distribution of unequal module sizes in previous work~\cite{sarkar2013a, sarkar2013b}, where the eigenvalues formed clusters based on the distributions of module sizes. Here, results are presented for all communities of the same size, since we also explore the rotation behaviour up to the theoretical detectability limit.  

\subsection{The theoretical detectability limit}

In this section we discuss the behaviour of eigenvalues when, keeping probability of connections inside a module or community constant ($\mu_{in}$, $p_{in}$, or $c_{in}$), we increase the probability of connections between modules (i.e., steadily increase $\mu_{out}$, $p_{out}$, or $c_{out}$). 

Now, we can write $A = \bar{A} + X$, where $\bar{A}$, the ensemble average matrix, is also partitioned into $q^{2}$ blocks of size $s$, with $q$ diagonal blocks with all entries equal to $p_{in}$, and $q(q-1)$ off-diagonal blocks with all entries equal to $p_{out}$. The fluctuations around the average $X$, by definition, has mean 0 and finite variance. As mentioned before, the eigenvalues of $A$ are denoted by $\{z_{1}\geq z_{2}\geq \ldots\geq z_{N}\}$. Let us denote the eigenvalues of $\bar{A}$ by $\{\lambda_{1} \geq\lambda_{2}\geq\ldots\geq\lambda_{N}\}$ and the eigenvalues of $X$ by $\{x_{1}\geq x_{2}\geq \ldots\geq x_{N}\}$. 

Since $A$, $\bar{A}$, and $X$ are all symmetric, we use the Weyl inequalities which imply that spectrum of a symmetric (Hermitian in the general case) matrix will be stable to small perturbations~\cite{taoBlog_254A_3a}: 
\begin{equation}
\label{weyl0}
z_{i+j-1} \leq \lambda_{i} + x_{j}, 
\end{equation} with $i,j \geq 1$ and $i + j -1 \leq N$. We are particularly interested in the following cases, that give us lower and upper bound estimates for $z_{1}$ and $z_{2:q}$:
\begin{equation}
\label{weyla}
z_{1} \leq \lambda_{1} + x_{1}, i,j=1,
\end{equation}
\begin{equation}
\label{weylb}
z_{2:q} \leq \lambda_{2:q} + x_{1}, i=2:q,j=1.
\end{equation}
This shows that when there are no fluctuations around the mean, then $z_{1} = \lambda_{1}$, and $z_{2:q} = \lambda_{2:q}$, but as fluctuations around the mean increase, $z_{1}$ and $z_{2:q}$ increase, but are still bounded by the inequalities~[\ref{weyla}] and [\ref{weylb}]. Thus, the lower bound on $z_{i}$ of $A$ is $b_{l} = \lambda_{i}$, the upper bound is $b_{u} = \lambda_{i} + x_{1}$. 

Thus, the actual $z_{i}$ lie anywhere between these upper and lower bounds. This implies an argument for detecting the modules, that is when the mean of these bounds $b = (b_{l} + b_{u})/2 = (\lambda_{i} + \lambda_{i} + x_{1})/2 = \lambda_{i} + (x_{1}/2)$ is subsumed into the bulk distribution, and is equal to $x_{1}$, it will no longer be possible to detect all the modules (as it is to be reasonably expected that at least one or some of the eigenvalues would have passed into the bulk by the time the mean passes into the bulk). As we will see, this provides us with the generalised theoretical detectability limit, that was proved for 2 communities in~\cite{nadakuditi2012}.  

Now we compute the eigenvalues of $\bar{A}$ and $X$. The eigenvalues of $\bar{A}$ can be easily calculated~\cite{sarkar2013a}. The first eigenvalue is the largest, with 
\begin{align}
\label{eq5} 
\lambda_{1} =s[p_{in} + (q-1)p_{out}], \\
\lambda_{1} = \frac{1}{q}[c_{in} + (q-1)c_{out}],
\end{align}
where $\lambda_{1}$ can also be expressed in terms of $c_{in}$ and $c_{out}$, with $N \to \infty$. Similarly, the next $q-1$ are
\begin{align} 
\label{eq6} 
\lambda_{2:q} =s(p_{in} - p_{out}), \\
\lambda_{2:q} =\frac{1}{q}[c_{in} - c_{out}],
\ 
\end{align} 
where $\lambda_{2:q}$ implies eigenvalues from $2$ to $q$, while the remaining eigenvalues are zero{
\begin{equation} \label{eq6a} 
\lambda_{q+1:N} =0.
\end{equation} 

We now derive the eigenvalue distribution for the matrix $X$, which is symmetric, has a mean of 0, and finite variance. By Wigner's semicircle law~\cite{furedi1981}, all the eigenvalues of $X$ will be contained in the interval $[-2\sigma_{A} \sqrt{N}, 2\sigma_{A} \sqrt{N}]$, where $\sigma_{A}$ represents the standard deviation of entries in the matrix $A$. For our case, this  implies that the eigenvalues of $X$ are spread around 0 but its largest one is $2\sigma_{A}\sqrt{N}$. 

We now derive the variance $\sigma^{2}_{A}$.
 $A$ has only 0 and 1 entries. The mean expected value of $A$ is $M = \frac{1}{q}[p_{in} + (q-1)p_{out}]$, thus the number of 1 entries, columnwise, is $NM$, and the number of 0 entries, columnwise, is 
$N(1 -M)$. Calculating variance by its definition, 
\begin{flalign}
\sigma^{2}_{A} &= [NM(1 - M)^{2} + N(1 - M)(0 - M)^{2}]/N,& \\
&= M(1-M)^{2} + M^{2}(1-M),& \\
&= M(1-M),& \\
\label{eq12a}
&= \frac{1}{q}[p_{in} + (q-1)p_{out}] - (\frac{1}{q}[p_{in} + (q-1)p_{out}])^{2}.
\end{flalign}
The variance for each column is the same, so, the variance for the whole of $A$ is as shown in Eq.~[\ref{eq12a}]. Further, as $N$ grows, $M(1-M) \approx M$, thus we can say $\sigma_{A} \approx \sqrt{M}$. 

Now, following~\cite{furedi1981}, the largest eigenvalue of $X$ can be computed using $2\sigma_{A} \sqrt{N}$ as
\begin{equation}
x_{1} = 2\sqrt{NM} = 2\sqrt{\lambda_{1}}.
\end{equation}
Using Eqs~(\ref{weyla}) and (\ref{weylb}), the $z_{2:q}$ fall between the lower and upper bounds of $b_{l} = \lambda_{2:q}$ and $b_{u} = \lambda_{2:q} + x_{1}$, with $b = \lambda_{2:q} + (x_{1}/2)$ . Since $\lambda_{2:q}$ provide the lower limit, one threshold is attained when
\begin{equation}
\lambda_{2:q} = 2\sqrt{\lambda_{1}}. 
\end{equation}
This provides the lower threshold [demonstrated in Fig.~\ref{Fig1}(b-e)]
\begin{flalign}
&\lambda_{2:q} > 2\sqrt{\lambda_{1}},& \\
&s(p_{in} - p_{out}) > 2 \sqrt{s(p_{in} + (q-1)p_{out}},& \\
&qs(p_{in} - p_{out}) > 2 \sqrt{q^{2}s(p_{in} + (q-1)p_{out}},& \\
&c_{in} - c_{out} > 2 \sqrt{q[c_{in} + (q-1)c_{out}]}.
\end{flalign}  
This also implies the condition $\lambda_{1} > \lambda_{2:q} > 2\sqrt{\lambda_{1}}$. Since $\lambda_{2}$ is the lowest possible value of $z_{2}$, this is the lowest limit, that is, it is ensured that if this condition is satisfied, then the modules will be detected absolutely. This threshold marks the beginning of the ``hard" phase, where even though the actual $z_{2:q}$ still sit outside the bulk, it gets progressively harder to detect them as $c_{out}$ increases further, see Fig.~\ref{Fig1}(b-e) for demonstration. 

Further, let us now consider the eigenvalues $z_{2:q}$: when these eigenvalues move into the bulk, it will provide the upper bound for the detectability, since once all or some of them move into the bulk, it will no longer be possible to detect the communities. As already mentioned, since the eigenvalues $z_{i}$ are bounded by $b_{l}$ and $b_{u}$, a condition for not detecting all the modules is when $b$ becomes equal to the largest eigenvalue of the bulk distribution, $x_{1}$. When we set $b$ equal to the largest eigenvalue of the bulk $x_{1}$, as before, we get
\begin{flalign}
\frac{1}{2} [2\lambda_{2} + 2\sqrt{\lambda_{1}}] &= 2\sqrt{\lambda_{1}},& \\
\lambda_{2} &= \sqrt{\lambda_{1}},
\end{flalign}
which will give us the general detectability limit for $q$ communities, as stated in previous works and derived for $q=2$~\cite{nadakuditi2012, krzakala2013}, and demonstrated in Fig.~\ref{Fig1}(b-e), 
\begin{equation}
c_{in} - c_{out} = \sqrt{q[c_{in} + (q-1)c_{out}]}.
\end{equation}

This ``hard" phase also shows that it will be hard to detect the modules even before the absolute detectability limit. For the stochastic block model demonstrated in Fig.~\ref{Fig1}, the actual $z_{i}$ are smaller than $b$ and as the value of $c_{out}$ is increased, it becomes impossible to detect the modules or their number even before the threshold is reached. For example, at a $c_{out}$ value of 3.5, which is just below the detectability threshold (3.9), it is already not possible to clearly detect the modules, Fig.~\ref{Fig1}(d). This behaviour has also been empirically reported in~\cite{krzakala2013}. With the analytic insight provided here, we establish that the detectability properties begin to deteriorate in a ``hard" phase of detection that is characterised by the bounds. 

For theoretical interest, if we push the same idea to its other extreme limit, that is the point where the upper limit $\lambda_{2} + x_{1}$ become equal to $x_{1}$, we will see that $\lambda_{2}$ goes to zero: this implies no modularity in the system, only the bulk. If $c_{out}$ increases even further, we can hypothesise that the $\langle z_{2:q} \rangle$ will move towards and out of the other end of the bulk, and the groups would be distinguishable again. They would not be communities, though, but anti-communities, as they will be much more connected with other groups than they are internally.

The general detectability limit provides the condition on the gaps between eigenvalues that governs the perturbation behaviour of the eigenvectors, used to detect the numbers and compositions of modules, as shown in the next section. 

\subsection{Detecting number of communities through rotation of eigenvectors under perturbation}

The first eigenvalue of $\bar{A}$ is the largest, 
$
\label{eq5}
\lambda_{1} =s[p_{in} + (q-1)p_{out}]$, 
the next $q-1$ are $ 
\label{eq6} 
\lambda_{2:q}=s(p_{in} - p_{out})$, where $\lambda_{2:q}$ implies all eigenvalues from $2$ to $q$,  
while the remaining eigenvalues are $\label{eq6a} 
\lambda_{q+1:N} =0$. 
Of particular interest here are the following gaps, where we consider the limit $b$ as an estimate for the value $\langle z_{i} \rangle$:  
\begin{flalign} 
\label{eq7} 
\delta_{1} &=\lambda_{1}-\lambda_{2:q}=Np_{out},& \\
\label{eq7a}
\Delta_{1} &= \langle z_{1} \rangle - \langle z_{2:q} \rangle = \delta_{1},& \\
 \label{eq8} 
\delta_{2} &= \lambda_{2:q} - 2\sqrt{\lambda_{1}},& \\
\label{eq8a}
\Delta_{2} &= \langle z_{2:q} \rangle - 2\sqrt{\lambda_{1}} = \lambda_{2:q}  - \sqrt{\lambda_{1}},& \\
\label{eq9} 
\delta_{3} &= 2\sqrt{\lambda_{1}} - \lambda_{q+1:N} = 2 \sqrt{\lambda_{1}}. 
\end{flalign} 
We now perturb $A$ with $\epsilon E$, getting $A + \epsilon E = (\bar{A} + X) + \epsilon E = \bar{A} + X'$, where once again $X'$, by definition, has mean 0 and finite variance. Thus, $\bar{A}$ is a rank $q$ matrix, with $q<N$. We then substitute $\langle z_{i} \rangle$ and $\Delta_{i}$ into the 
new improved bounds [Eqs(\ref{eq3}) and (\ref{eq4b})].

First, for $v_{1}$, substituting the values of $\langle z_{1} \rangle$ and $\Delta_{1}$ into Eq.~(\ref{eq3}), if we fix $q$, $p_{in}$, and $p_{out}$, all three terms decrease as $N$ grows. 
Thus, the rotation of the first eigenvector is bounded to a small angle, as seen in Fig.~\ref{Fig1}(a). 


Second, eigenvectors $2 \ldots q$ span a subspace and rotate together. Defining the subspace $V = \{v_{2}, \ldots, v_{q}\}$ and $V' = \{v'_{2}, \ldots, v'_{q}\}$, the largest principal angle between all pairs of vectors is governed by $\Delta_{2}$. Substituting the values of $\langle z_{2:q} \rangle$ and $\Delta_{2}$ into Eq.~(\ref{eq4b}), with $q$, $p_{in}$, and $p_{out}$ fixed, again implies that all three terms decrease as $N$ grows. Thus, the rotation of eigenvectors $2,\ldots,q$ are also bounded to a small angle, Fig.~\ref{Fig1}(a). Since $\Delta_{1}$ and $\langle z_{1} \rangle$ are larger than $\Delta_{2}$ and $\langle z_{2} \rangle$, the sine of the angle between $V$ and $V'$ is larger than that between $v_{1}$ and $v'_{1}$, 
but still bounded to a small angle with high probability governed by $\Delta_{2}$. Figure~\ref{Fig1}(a) shows that eigenvectors $v_{2:q}$ indeed have this behaviour: they rotate as a group, showing that the subspace $V$ behaves as one, and is different from the subspaces $v_{1}$ and $V'' = \{v_{q+1}, \ldots, v_{n}\}$. We call $V$ the \textit{community eigenspace} and $V''$ the \textit{bulk eigenspace}, for which applying the same theorems will lead to the largest angles of rotation with the sine approaching $90^{\circ}$.   
The results in Fig.~\ref{Fig1}(a) show a clear sharp separation between the community eigenspace and the bulk eigenspace, revealing the correct number of modules in the network.
This behaviour changes as we approach the detectability limit for sparse modular networks, a threshold beyond which modularity exists in the network, but cannot be detected.

Figure~\ref{Fig1}(b)-(e) shows how the zone between $\lambda_{2:q}$ and $\lambda_{2:q} + x_{1}/2$ gradually moves into the bulk as $c_{out}$ or $\mu_{out}$ are increased, keeping $c_{in}$ or $\mu_{in}$ constant. $\Delta_{2}$ gradually decreases as $\lambda_{2:q} + x_{1}/2$ moves towards the bulk, and becomes equal to $x_{1}$, providing the detectability threshold. 

Figure~\ref{DifferencePlot} shows the differences between the mean angles of rotation of the first eigenvector of the bulk [the $(q+1)^{th}$, though choosing any vector for this plot would still reveal $q$ due the structure in Fig.~\ref{Fig1}(a)] and of all the other eigenvectors of $A$, $D_{q+1} - D_{i}, i = 2:N$, for $\mu_{out}=0$ to $10$ keeping $\mu_{in} = 10$. This difference brings out the behaviour of the two subspaces clearly: for eigenvectors $2$ to $q$, $D_{q+1} - D_{i}$ decreases as $\mu_{out}$ is increased, whereas for the eigenvectors of the bulk the behaviour is different. 

At the detectability threshold (vertical line), this difference is the smallest for all the vectors, after the threshold it starts to expand again, for both the community and the bulk eigenvectors. The larger this difference, the clearer the separation between the community and the bulk eigenspaces, and the easier it is to detect the number of groups. 

Note that once the ``hard" phase of detection sets in, there is a steep drop and a near zero gap to the detectability limit, even before the exact limit is reached. At the detectability threshold, the rotation angles of the community and the bulk eigenspace merge into one. If $\mu_{out}$ increases further, we can hypothesise that the $\langle z_{2:q} \rangle$ will move towards and out of the other end of the bulk, and the groups would be distinguishable again. They would not be communities, though, but anti-communities, as they will be much more connected with other groups than they are internally.  
\begin{figure}
\includegraphics[width = 0.45\textwidth]{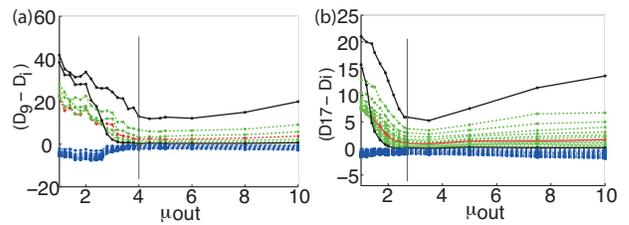}
\caption{\label{DifferencePlot} [Color online]. Difference between mean angles of rotation of the $(q+1)^{th}$ and of the top 50 eigenvectors of stochastic block models under perturbations, $D_{q+1} - D_{i}, i = 2:50$ as $\mu_{out}$ is varied from $0$ to $10$, keeping $\mu_{in} = 10$. (a) $N=4096$,  $q = 8$ and detectability threshold $\mu_{out}^D= 4$. (b) $N=10000$, $q = 16$ and detectability threshold $\mu_{out}^D= 2.7$. Each point is an average over $10$ networks, each perturbed $300$ times. Rotation lines for eigenvectors $2^{nd}$ and $q^{th}$ in black, all other community eigenvectors in green, the middle one between the $2^{nd}$ and the $q^{th}$ in red, bulk eigenvectors in blue.}
\end{figure}  

In addition to the above results, we also empirically observe an oscillatory behaviour that is not explained by the theorems above: the pairing up of eigenvectors corresponding to ``mirrored" eigenvalues in both the community and bulk eigenspaces. For example, in Fig.~\ref{Fig1}(a), with 16 modules, the angles of rotation under perturbation for the first 16 eigenvectors are separated from the bulk. The angles of rotation of the community eigenvectors, e.g., the $2^{nd}$ and the $16^{th}$, the $3^{rd}$ and the $15^{th}$, and so on, and those of the bulk, e.g. the $17^{th}$ and the $10000^{th}$, and so on, are similar. We observed this behaviour across a large range of parameters and networks. The angles of rotation first increase for the first half of the eigenvectors, and then decrease again for the last half of the eigenvectors in the subspace, resulting in a symmetric pairing up of eigenvectors from the two halves.

Figure~\ref{Fig3}(a) shows the distribution of eigenvalues in the community and bulk eigenspaces along with the mean eigenvalue of each eigenspace.
\begin{figure}
\includegraphics[scale=0.2]{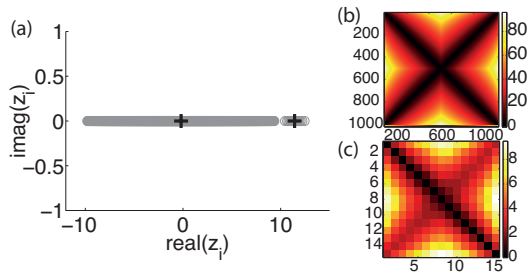}
\caption{\label{Fig3} [Color online] Symmetric distribution of eigenvalues as basis for ``pairing up" of eigenvectors under rotation. (a) Eigenvalue distribution for a network generated by a stochastic block model with $N = 10000$, $q=16$, $\mu_{in} = 10$, $\mu_{out} = 1$. Black markers show mean eigenvalues for the community and bulk eigenspaces. (b),(c). Distance matrix $\bf{W}$ for bulk and community eigenspaces, respectively.}
\end{figure}
We characterise the distribution of eigenvalues in the two eigenspace distributions by defining deviations of each eigenvalue from the mean of the eigenvalues in each space. These are defined as $w_{i}=| z_{2:q}-\langle z_{2:q} \rangle|, i = 2:q$, and $w_{i}=|z_{q+1:N}-\langle z_{q+1:N} \rangle|, i = q+1:N$.  
Over $m$ networks, we get vectors ${\bf w}_{i} \in \Re^{m}$, $i = 1:N$, where each ${\bf w}_{i}$ is a measure of the deviation from the mean eigenvalues for the groups defined above. We compute a distance matrix $W$ with $W_{ij}$ equal to the Euclidean distance between the vectors ${\bf w}_i$ and ${\bf w}_{j}$, with $i,j = 1$ to $N$. Figures~\ref{Fig3}(b) and (c) show $W$ for the bulk and the community spaces, respectively. The main diagonals in both show that the eigenvalues successively close to each other are at very low distance, but the main cross-diagonal shows the same low-distance relationship for pairs of eigenvalues symmetrically disposed about the mean eigenvalue of the eigenspace, showing self-similar behaviour in both the bulk and in the community eigenspace; i.e., Fig.~\ref{Fig3}(c) is simply a blow up of the bottom-right corner of Fig.~\ref{Fig3}(b). Since the gaps between eigenvalues govern the rotation behaviour, we empirically relate observations on the distributions of eigenvalues, gaps between them, and their symmetric distribution around the means: not only the rotation angles depend on the gaps, but the pairing of eigenvectors in each eigenspace seems to be related to the pairing of eigenvalues via their symmetric distributions around the mean eigenvalue in that eigenspace. This behaviour is lost at the detectability limit, where the two spaces merge to become one, as shown both by the eigenvalue gaps [Fig.~\ref{Fig1}(b-e)] and the rotation of eigenvectors under perturbation (Fig.~\ref{DifferencePlot}).   

\subsection{Tests on real and benchmark networks}

Finally, the SBM, while very useful for deriving analytical results, does not accurately represent the structure of real world networks. Therefore, we tested the approach on Lancichinetti-Fortunato-Radicchi (LFR) benchmark graphs~\cite{lancichinetti2008} that can be used to generate networks that have some properties of real world networks such as broad degree and community size distributions and benchmark real world networks with known community structure. We generated a number of these networks, parametrically defining a complex mix of parameters such as a high number of very small communities with differing sizes and varying degree. 

In Fig.~ [\ref{Fig_benchmarkReal}](a) and (b), we use LFR networks with 1000 nodes and 24 and 43 communities, respectively, and vary the mixing parameter $\mu$, expressing the ratio between the external degree of a node and its total degree. The number of communities is correctly estimated by the rotation of eigenvectors under perturbation. 

The plot in Fig. 4(a) shows the community and bulk eigenspaces, respectively, with the community eigenspace defined by the first 24 eigenvectors corresponding to the 24 largest eigenvalues, respectively, and the bulk eigenspace defined by the rest. We note that not only is there a sharp separation gap between the $24^{th}$ and $25^{th}$ eigenvectors, there is also the corresponding oscillatory behaviour in the community and bulk eigenspaces, revealing exactly the 24 communities. Exactly the same reading holds for Fig. 4(b), showing an LFR network with 43 communities. These communities can then be detected by using any algorithm, but most obviously, by using a lower dimensional space defined by 28 (or 43) eigenvectors in each case and defining each vertex as a point in 28 or 43-dimensional space and using cosine or K-means clustering to detect the modules accurately~\cite{sarkar2011a}. The gap is less prominent in Figure 4(b) because $\mu$ is higher and communities are more mixed and harder to detect.  

We also apply the approach to some real world benchmark networks. Figure 4(c) and (d) show the American college football network~\cite{girvan2002} and the network of political books~\cite{krebs06}. The American College Football network~\cite{girvan2002} is a network of American football games between college teams during a regular season (Fall 2000). Teams are organised into conferences, with each conference containing around 8-12 teams. Games are more frequent between teams of the same conferences than between teams of different conferences. This results in communities. The network of political books~\cite{krebs06} is a data incorporating books about recent US politics sold by the online bookseller Amazon.com. Edges between books represent frequent co-purchasing of books by the same buyers. The network was compiled by V. Krebs and is unpublished, but can found on Krebs' web site. Two main communities exist, representing two main political party divisions, since members supporting one party are more likely to purchase books representing that party's ideology. In both cases, the correct number of communities is predicted, and lower dimensional spectral detection can similarly be employed to detect the communities accurately.

\begin{figure}
\includegraphics[scale=0.19]{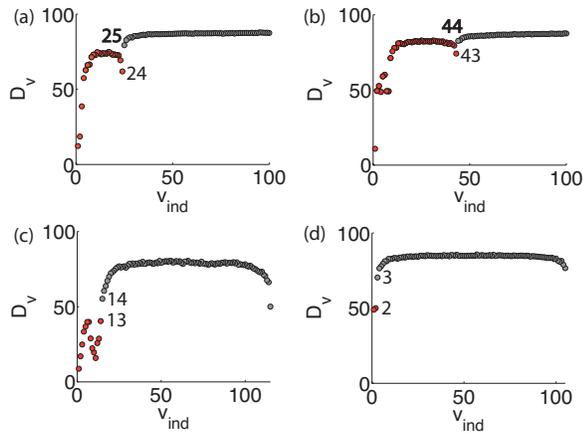}
\caption{\label{Fig_benchmarkReal} [Color online] Number of communities (red circles) detected in LFR and real networks by eigenvector rotations. Each network is perturbed 300 times, the eigenvector index $v_{ind}$ is plotted against its angle of rotation $D_{v}$. (a) LFR network with $N=1000$, $q=24$, $\langle k\rangle= 25$, $k_{max}= 60$, $\mu= 0.3$. (b) LFR network with $N=1000$, $q=43$, $\langle k\rangle= 25$, $k_{max}= 60$, $\mu= 0.4$. (c) American college football network: $q=13$, $N=115$~\cite{girvan2002}. (d) Political books network: $q=2$, $N=105$~\cite{krebs06} (see Appendix).}
\end{figure}

\section{Conclusions and Discussion}
We presented the dynamics of rotation of eigenvectors of adjacency matrices of modular networks under random perturbations. In the presence of $q$ communities, the number of eigenvectors corresponding to the $q$ largest eigenvalues form a ``community" eigenspace and rotate together, but separately from that of the ``bulk" eigenspace spanned by all the other eigenvectors. Using this property, the number of modules or clusters in a network can be accurately estimated in an algorithm-independent way. Results are shown to hold to a point where a ``hard" phase of detectability sets in before the theoretical detectability limit for sparse modular networks. Analytic insight is presented into the bounds of this hard phase, using which a general derivation of the detectability threshold is presented for $q$ communities, using arguments based on these bounds. This is consistent with previous results, and a proof of which was previously provided for 2 communities. A plausibility argument is presented for the observed symmetric pairing up of eigenvalues and eigenvectors in the two eigenspaces. The approach presented demonstrates that the rotation behaviour of the eigenvectors of the adjacency matrix under perturbations reveals information about the community structure of a network.   
\vspace{1cm}

\begin{acknowledgments}
This work is supported by a Henry Halloran Trust Research Fellowship, University of Sydney, Australian Research Council (ARC) Center of Excellence for Integrative Brain Function Grant CE140100007, ARC Laureate Fellowship Grant FL1401000225, and ARC Discovery Project Grant DP130100437. 
\end{acknowledgments}

\end{document}